\documentclass[preprint,proceedings]{rmaa}

\usepackage{paralist}

\usepackage{psfrag,color}

\SetYear{2005} \SetConfTitle{Triggering of Relativistic Jets}

\title{Launching of Jets by Propeller Mechanism}

\author{
  M. M. Romanova,\altaffilmark{1}
  R. V. E. Lovelace,\altaffilmark{1}
  G. V. Ustyugova,\altaffilmark{2}
  and A.V. Koldoba, \altaffilmark{2,3}}

\altaffiltext{1}{Cornell University, Ithaca, NY USA.}
\altaffiltext{2}{Keldysh Institute of Applied Mathematics, Moscow,
Russia .} \altaffiltext{3}{Institute of Mathematical Modelling,
Moscow, Russia.}

\shortauthor{Romanova, et al.}
\shorttitle{RevMexAA(SC) Propeller Outflows}

\abstract{We carried out axisymmetric simulations of disk accretion
to a
 rapidly rotating magnetized star in the ``propeller" regime.
 Simulations show that propellers may be ``weak" (with no
 outflows), and ``strong" (with outflows). Investigation of the
 difference between these two regimes have shown that outflows
 appear only in the case where the ``friction"
 between the disk and  magnetosphere is sufficiently
 large, and when accreting matter flux is not very small.
   Matter outflows  in a wide cone and is magneto-centrifugally 
   ejected from the inner regions of the disk.
     Closer to the axis there is
a strong, collimated, magnetically dominated  outflow of energy and
angular momentum carried by the  open magnetic field lines from the
star.
  The ``efficiency'' of the propeller may be
very high in the respect that most of the incoming disk matter is
expelled from the system in winds.
   The star spins-down rapidly
due to the magnetic interaction with the disk through closed field
lines and with corona through  open field lines.
      This mechanism may act in a
variety of situations where magnetized star rotates with
super-Keplerian velocity at the magnetospheric boundary. We
speculate that in general any object rotating with super-Keplerian
velocity may drive outflows from accreting disk, if the friction
between them is sufficiently large.}

\addkeyword{ISM: Jets and outflows} \addkeyword{Stars: magnetic
fields}

\begin{document}
\maketitle

\section{Introduction}
\label{sec:intro}

Fast rotating accreting magnetized neutron stars or white dwarfs are
expected to be in the propeller regime during their evolution (
Davidson \& Ostriker 1973; Illarionov \& Sunyaev 1975; Stella,
White, \& Rosner 1986; Lipunov 1992; Treves, Colpi \& Lipunov 1993;
Cui 1997; Alpar 2001; Mori \& Ruderman 2003).
    The propeller regime is characterized by the fact that
the azimuthal velocity of the star's outer magnetosphere is larger
than the Keplerian velocity of the disk at that distance.

Different aspects of the propeller regime  were investigated
analytically (Davies, Fabian \& Pringle 1979; Li \& Wickramasinghe
1997; Lovelace, Romanova \& Bisnovatyi-Kogan 1999; Ikhsanov 2002;
Rappaport, Fregeau, \& Spruit 2004; Eksi, Hernquist, \& Narayan
2005) and studied with computer simulations. However only the case
of quasi-spherical accretion was investigated in the axisymmetric
simulations (Romanova et al. 2003) and the instabilities in the
equatorial plane were investigated in case of disk accretion in 2D
simulations (Wang \& Robertson 1985).

\begin{figure}[!t]
  \includegraphics[width=\columnwidth]{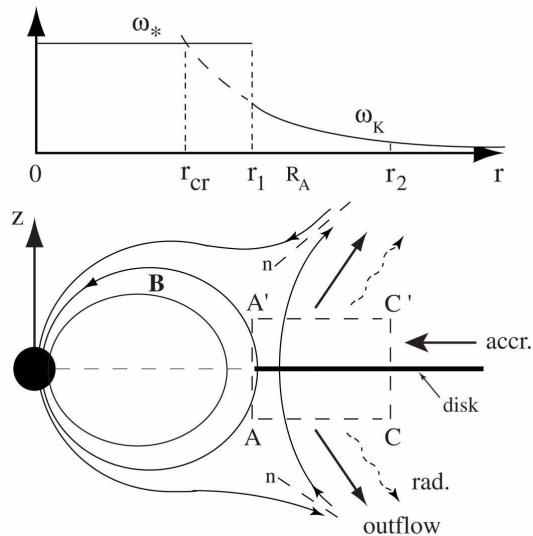}
  \caption{The sketch demonstrates the basic principles of the propeller
  stage (from Lovelace et al. 1999)}
  \label{fig:simple}
\end{figure}

We performed axisymmetric (2D) simulations of disk accretion to a
fast rotating star in the propeller regime. Simulations have shown
that in some cases a star spins-down but no outflow form (Romanova
et al. 2004, - hereafter RUKL04). In other cases strong outflows
form (Romanova et al. 2005 - hereafter - RUKL05); Ustyugova et al.
2005). In these Proceedings we discuss launching of jets and winds
by propeller mechanism

\section {A model}

We have done axisymmetric MHD simulations of the interaction of an
accretion disk with magnetosphere of a rapidly rotating star.
      What
is meant by rapid rotation is that the corotation radius of the star
$r_{cr}=(GM_*/\Omega_*^2)^{1/3}$ is smaller than the magnetospheric
radius $r_m$ which is determined by the balance between the pressure
of the star's magnetic field and the ram pressure of the disk
matter, that is where modified plasma parameter $\beta = (p + \rho
{\bf v}^2)/({\bf B}^2/8\pi) = 1$. Here ${\bf B}$ is the surface
magnetic field of the star, $\Omega_*$ is angular velocity of the
star.

The numerical model we use
  is similar to that of Romanova et al. (2002 - hereafter - RUKL02)
and  RUKL04).
   Specifically, (1) a spherical
coordinate system (r,$\theta$,$\phi$) is used  to give high
resolution near the dipole; (2) the complete set of MHD equations is
solved to find the eight variables ($\rho, v_r, v_\theta, v_\phi,
B_r, B_\theta, B_\phi, \varepsilon$) (with $\varepsilon$ the
specific internal energy); (3) a Godunov-type numerical method is
used;
   (4) special ``quiescent" initial
conditions were used so that we were able to observe slow viscous
accretion from beginning of simulations (see details in RUKL02).

   We suggest that both viscosity
and diffusivity are determined by turbulent fluctuations of the
velocity and magnetic field (e.g., Bisnovatyi-Kogan \& Ruzmaikin
1976) with both the kinematic viscosity $\nu_{\rm t}$ and the
magnetic diffusivity $\eta_{\rm t}$ described by
$\alpha-$coefficients as in the Shakura and Sunyaev model.
      That is, we take
$\nu_{\rm t}=\alpha_{\rm v} c_s^2/\Omega_K$ and $\eta_{\rm
t}=\alpha_{\rm d} c_s^2/\Omega_K$, where $\Omega_K$ is the Keplerian
angular velocity in the disk, $c_s$ is the isothermal sound speed,
and $\alpha_{\rm v}$ and $\alpha_{\rm d}$ are dimensionless
coefficients $\lesssim 1$.
    In RUKL04 we investigated a range of small viscosities and
diffusivities, $\alpha_{\rm v},~\alpha_{\rm d}\sim 0.01-0.02$ and
found no significant matter outflows.
    This paper  investigates a
wider range of $\alpha-$ parameters and finds substantial outflows
for $\alpha_{\rm v} \gtrsim 0.1$ and $\alpha_{\rm d} \gtrsim 0.1$ in
the propeller regime.

\begin{figure}[!t]
  \includegraphics[width=\columnwidth]{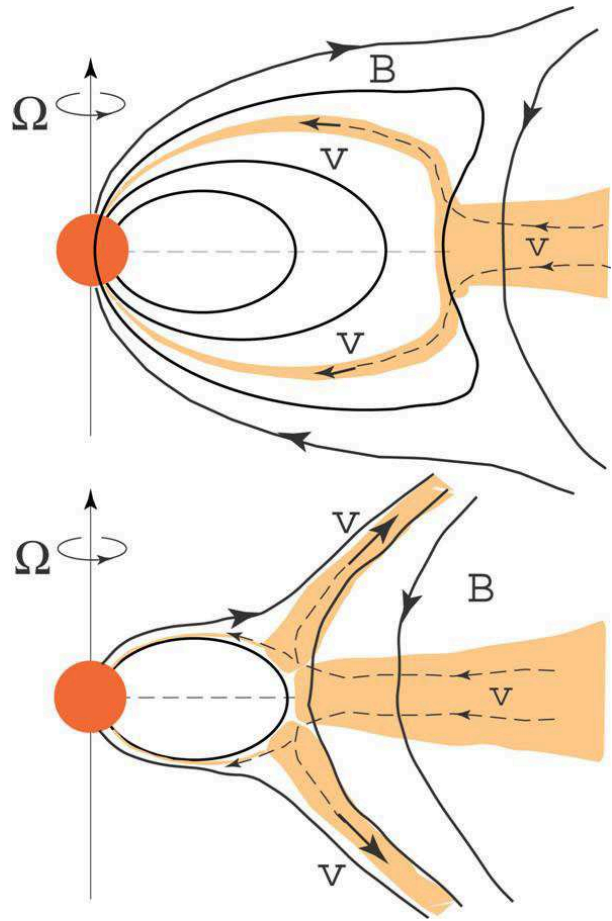}
  \caption{The sketch demonstrates the difference between the
  weak (top panel) and strong (bottom panel) propellers.}
  \label{fig:simple}
\end{figure}

\section{``Weak" and ``Strong" Propellers}

Simulations have shown that there are two types of propellers:
``weak" propellers, at which a star strongly spins-down, but no
significant outflows were observed (RUKL04), and ``strong"
propellers, at which a robust outflows of matter and magnetic energy
are launched from the magnetosphere of the propelling star (RUKL05,
Ustyugova et al. 2005). There are two conditions which are necessary
to launch the outflows:

\begin{asparaitem}

\item Relatively high ``friction" between magnetosphere and the disk,
which we determined by viscosity and diffusivity;

\item Relatively high specific matter flux (flux per unit area) in
the disk, which is regulated by a number of parameters, including
viscosity.

\end{asparaitem}

 Figure 2 demonstrates the difference between weak and
strong propellers. Top panel shows weak propeller, where matter flux
is not strong enough to penetrate inward and also the ``friction"
between magnetosphere and the disk is not sufficient to launch
outflows, and the bottom panel shows the case when outflows are
successfully launched.

\begin{figure}[!t]
  \includegraphics[width=\columnwidth]{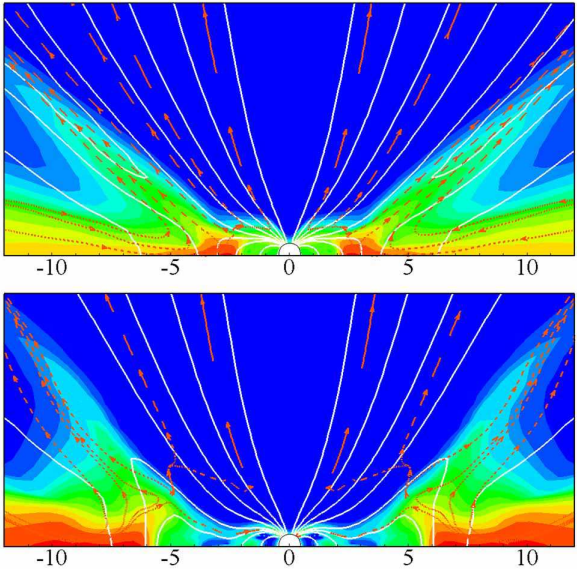}
  \caption{Matter flow in the
  ``high" (top panel) and ``low" (bottom panel) states. The background
  shows the density distribution, white solid lines are magnetic field
  lines, vectors are velocity vectors. The top panel shows
  results of simulations after $T=924$ Keplerian periods of rotation
   $P_0$ at $r=1$. The bottom panel shows the flow $10$
   periods later.}
  \label{fig:simple}
\end{figure}

\begin{figure}[!t]
  \includegraphics[width=\columnwidth]{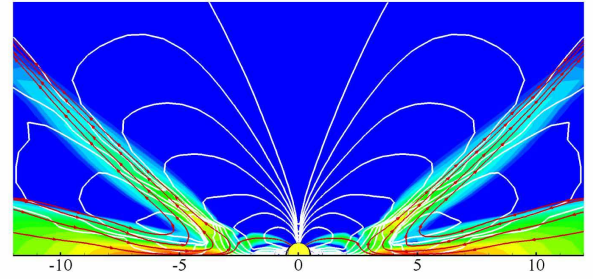}
  \caption{The background and streamlines with arrows show
  the distribution of angular momentum carried by matter. White solid lines
  show distribution of angular momentum carried by the field.}
  \label{fig:simple}
\end{figure}

\begin{figure}[!t]
  \includegraphics[width=\columnwidth]{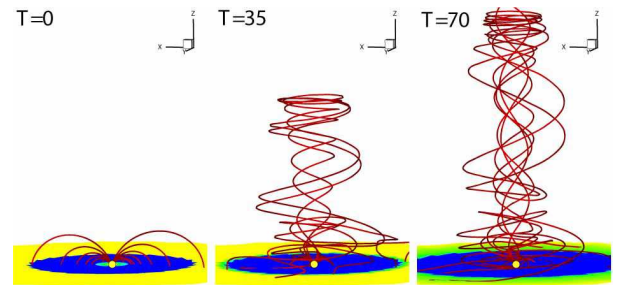}
  \caption{Magnetic field of the dipole strongly inflates as a result of the
  fast rotation of the star forming the magnetic ``tower" from RUKL04).}
  \label{fig:simple}
\end{figure}

\begin{figure}[!t]
  \includegraphics[width=\columnwidth]{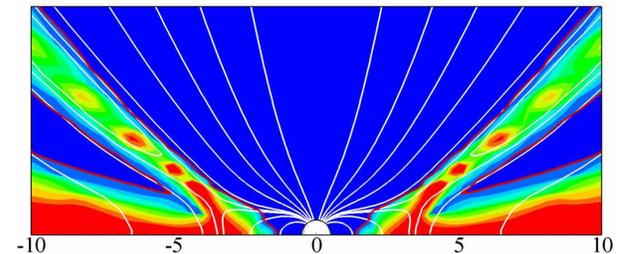}
  \caption{The background  shows distribution of the modified plasma
  parameter $\beta=(p+\rho v^2)/(B^2/8\pi)$ in the range from 1 to 100. Red solid 
  line corresponds to 
  $\beta=1$.}
  \label{fig:simple}
\end{figure}

In case of``weak" propellers (RUKL04), a star strongly spins-down,
but no significant outflows were observed. In case of  ``strong"
propellers significant part of the disk matter is re-directed to the
propeller-driven outflows (RUKL05, Ustyugova et al. 2005).

       We observed that the disk oscillates between
``high" and ``low" states and expels matter to  conical outflows
quasi-periodically. In the high state matter comes closer to the
disk and outflows are stronger. In the low state matter is pushed
back by expanding magnetosphere and outflows are weaker. Figure 3
shows matter flow in the high (top panel) and low (bottom panel)
states. Angular momentum of the disk is re-directed by the fast
rotating magnetosphere and flows to the conical wind (see Figure 4).
A star loses its angular momentum through magnetic stresses. Part of
this angular momentum is associated with the closed field lines
connecting a star with the inner regions of the disk. Other part
flows away along the open field lines connecting a star with corona.
 Fast rotation of the star at the propeller regime leads
to enhanced opening of the magnetic field lines, specifically those
located in the axial region. Field lines inflate and open forming
the magnetic ``tower" (see Figure 5). Most of the region is
magnetically-dominated excluding the disk and conical wind where
matter dominates. Figure 6 shows distribution of $\beta$. Most of
matter flows to this conical outflow, which we call the ``wind".
From other side, some matter flows to the axial region located above
the conical winds.  We call this part of outflow ``the jet". Jet
typically has lower density but much higher velocity. This is also
the region where significant energy flows due to the twisted
magnetic field lines (Poynting flux).

\section{Quasi-Periodic Oscillations}

Simulations have shown that the system oscillates between the high
and low states.
    The quasi-periodic outbursts associated with the disk-magnetosphere
interaction were discussed earlier by Aly \& Kuijpers (1990) and
observed in simulations by Goodson et al. (1997, 1999), Matt et al.
(2002), RUKL02, Kato et al. (2004), von Rekowski \& Brandenburg
(2004) and RUKL04. However, none of the earlier simulations
concentrated on the propeller stage, and only few oscillation
periods were obtained in earlier simulations. In recent simulations
of strong propellers numerous oscillations were observed (RUKL05).

In many cases the quasi-periodic oscillations are not very well
tuned (see Figure 7), and the time-scale of oscillations depends on
parameters. At larger magnetic moments $\mu$ and $\Omega_*$,
time-scale is longer. We also observed that at larger values of
viscosity oscillations become very well tuned (see Figure 8).

\begin{figure}[!t]
  \includegraphics[width=\columnwidth]{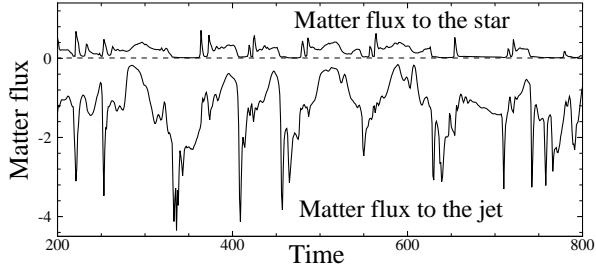}
  \caption{Variation of matter fluxes to the star and to the jet
  for the case $\alpha_{\rm v}=0.3$ and $\alpha_{\rm d}=0.2$.}
  \label{fig:simple}
\end{figure}

\begin{figure}[!t]
  \includegraphics[width=\columnwidth]{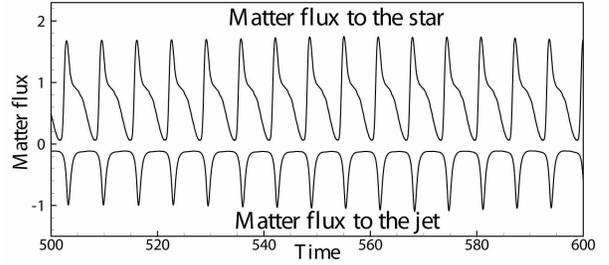}
  \caption{Variation of matter fluxes to the star and to the jet
  for the case $\alpha_{\rm v}=0.6$ and $\alpha_{\rm d}=0.2$
  (from RUKL05).}
  \label{fig:simple}
\end{figure}

\begin{figure}[!t]
  \includegraphics[width=\columnwidth]{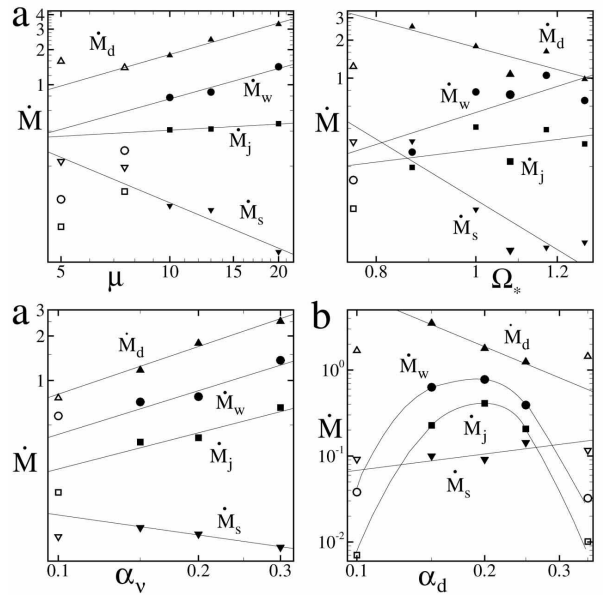}
  \caption{Dependence of matter fluxes on the main
  parameters of simulations for reference run. $\dot M_d$ - matter
  flux through the disk, $\dot M_w$ - to the wind, $\dot M_j$ -
  to the jet, $\dot M_s$ -
  to the star (from Ustyugova et al. 2005).} \label{fig:simple}
\end{figure}

\section{Angular Momentum Transport and  Spinning-down}
The angular
momentum flux carried by the disk matter is re-directed by the
rapidly rotating magnetosphere to the outflows (Figure 4).
      Furthermore, there is a strong outflow of angular
momentum carried by the twisted open magnetic field lines from the
star  $\langle N_f\rangle$, the Poynting flux and closed field lines
connecting a star and the disk.
    The time averaged total angular
momentum flux  from the star is ${\langle N_s\rangle}/{N_0} \approx
- 3.1 ({\mu_*}/{\mu_{00}})^{1.1} ({\Omega_*}/{\Omega_0})^{2.0}
({\alpha_{\rm d}}/{0.2})^{0.46} ({\alpha_{\rm v}}/{0.2})^{0.1}. $
  For our typical  parameters the
spin-down associated with open and closed field lines are
comparable.
   However, at larger $\Omega_*$ and/or $\bf{\mu}_*$, the
outflow along the open field lines dominates (see related cases in
Lovelace et al. 1995;
     Matt \& Pudritz 2004), while at
lower $\Omega_*$ or $\bf{\mu}_*$, the situation reverses and a
larger flux is associated with the closed field lines (like in Ghosh
\& Lamb 1979; RUKL02).

The  spin-down time-scale follows from derived value ${\langle
N_s\rangle}/{N_0}$ and the relation $N_*=I_* d\Omega_*/dt$, where
$I_*\approx 0.4 M_* R_*^2$ is moment of inertia of the star.
     For the period of the star
$P_*=2\pi/\Omega_*$, we obtain: $P_*(t) =P_*(0)(1+t/t_{\rm sd})$,
where the spin-down time is
$$ t_{\rm sd}\approx 0.036 \bigg(\frac{M_*}{\dot M_0}\bigg)
\bigg(\frac{\mu_{00}}{\mu_*}\bigg)^{1.1}
 \bigg(\frac{0.2}{\alpha_{\rm d}}\bigg)^{0.46}
\bigg(\frac{0.2}{\alpha_{\rm v}}\bigg)^{0.1} . $$

Our simulations may be directly applied only to stars with
relatively small magnetospheres, $r_m\sim 3-10 R_*$, this is why we
show below an example for millisecond pulsars.
    For a neutron star with
mass $M_*=1.4 M_\odot$ and accretion rate $\dot M_0\approx 5\times
10^{-8}~{\rm M_\odot/yr}$, we find the spin-down time
$t_{sd}\approx10^6~{\rm yr}$.
   Taking $R_*=10^6~{\rm cm}$ and
$\mu_*=\mu_{00}$, we find $\mu_*\approx7.0\times 10^{27}~{\rm G
cm^3}$, $B_*\approx7\times 10^9~\rm G$, $P_*\approx1.3~{\rm ms}$.
   In this case $t_{sd}$ represents the
time-scale of spin change for rapidly rotating millisecond pulsars.

One can see that for larger magnetospheres, the spin-down will be
faster.

\section{Efficiency of Propeller}

We changed the main parameters of the model: magnetic moment of the
star $\mu$, angular velocity of the star $\Omega_*$, viscosity
$\alpha_{\rm v}$ and diffusivity $\alpha_{\rm d}$ coefficients and
calculated matter fluxes to the wind $\dot M_w$, to the jet $\dot
M_j$, to the star $\dot M_s$ and total matter flux through the disk
$\dot M_d$. Figure 9 shows dependencies of these fluxes on the main
parameters. One can see that at larger magnetic moment and angular
velocity of the star, more matter flows to the winds/jets and less
matter accretes to the star. The efficiency of propeller may be very
high, so that most of disk matter may be re-directed to the jet and
wind. Matter fluxes to the wind and jet also increase with increase
of viscosity. Dependence on diffusivity, however, has a maximum near
$\alpha_{\rm d}\approx 0.2$. At smaller values, the diffusivity
helps to increase the ``friction" between magnetosphere and the
disk, because the disk matter  penetrates through the field lines of
magnetosphere and thus magnetosphere transports its angular
momentum. However at larger values of $\alpha_{\rm d}$ the slippage
of field lines through the disk matter become significant and the
coupling between the disk matter and magnetosphere decreases. It is
clear that in case of ``weak" propellers the coupling between
magnetosphere and the disk was not strong enough to transport the
angular momentum from the star to the disk, and no outflows were
observed.

\section{Possible Application to Black-Hole Systems}

It is interesting to note that disk oscillations and quasi-periodic
outbursts were observed in microquasars (see, e.g.,  these
Proceedings), which have a black hole in the center. The nature of
these oscillations is not known and the similarity between our
simulations and observations of microquasars may be accidental.
However, the question is discussed in the literature, that in case
of the black hole systems, the magnetic flux may accumulate in the
inner regions of the disk, and then reconnect, and this may lead to
quasi-periodic oscillations and outbursts to jets (e.g., Mayer,
these Proceedings). However, it is still not clear (1) how
magnetized matter of the inner regions of the disk interact with the
black hole, or (2) can rapidly rotating black hole influence to the
rotation of the inner regions of the disk through magnetic field
lines? If Kerr black hole transports its angular momentum to the
inner regions of the disk, then these regions of the disk will
rotate with super-Keplerian velocity and may act as propeller. In
fact, a star with the dipole magnetic field, or, a black hole may be
compared to the rotating super-Keplerian ball which transports part
of its angular momentum to the surrounding matter which may flow to
the jets during periods of enhanced accretion. Note, that in
microquasars periods of outflow are associated with enhanced
accretion rates (e.g., Novak, these Proceedings). Interesting to
note, that in full GR simulations of the disk accretion to a black
hole, it was observed that the black hole does transport angular
momentum to the disk (Hawly and Krolik, these Proceedings). Future
simulations are needed with the larger magnetic field in the disk to
understand the interaction of the rapidly rotating black hole with
the disk.

\section{Conclusions}

    In the propeller regime of disk
accretion to a rapidly rotating star, we find from axisymmetric MHD
simulations that the disk oscillates strongly and gives
quasi-periodic  outflows of matter to wide-angle ($\chi\approx
45^\circ - 60^\circ$) conical winds.
    At the same time there is strong field-dominated (or Poynting)
outflow of energy and angular momentum along the open field lines
extending from the poles of the star.
     The outflows occur for conditions where
the magnetic diffusivity and viscosity are significant, $\alpha_{\rm
v,d}\gtrsim 0.1$.
       For smaller values of the diffusivity, the disk
oscillates but no outflows are observed (RUKL04).
   The observed
oscillations and outbursts are a robust result, based on a numerous
simulations at different parameters with more than a $100$
oscillation periods observed in many runs.
  The period of oscillations varies in different runs in
the range $\tau_{\rm qpo}\sim (5-100) P_*$.
  It increases with
$\mu_*$ and $\Omega_*$.
    We observed that the oscillations for
relatively large $\alpha_{\rm v}$ become highly periodic with
definite quasi-periods.
    More detailed analysis of
these features will be reported later. A  star spins-down rapidly
due to both the disk-magnetosphere interaction and the angular
momentum outflow along the open field lines.

\acknowledgements

We thank the organizers for wonderful meeting. This work was
supported in part by NASA grants NAG5-13220, NAG5-13060, NNG05GG77G
and by NSF grants AST-0307817, AST-0507760. AVK and GVU were
partially supported by RFBR 03-02-16548 grant. The authors thank Dr.
Daniel Proga and Dave Rothstein for stimulating discussions.

\end{document}